\documentclass[prl,aps,twocolumn,superscriptaddress,showpacs,floatfix]{revtex4}
\usepackage{amsmath}
\usepackage{graphicx}

\def\cg#1#2{{\left\langle#1\vert#2\right\rangle}}
\def\threej(#1,#2)(#3,#4)(#5,#6){\begin{pmatrix}#1&#3&#5\\#2&#4&#6\end{pmatrix}}
\def\sixj(#1,#2,#3)(#4,#5,#6){\begin{Bmatrix}#1&#2&#3\\#4&#5&#6\end{Bmatrix}}
\def\ninej(#1,#2,#3)(#4,#5,#6)(#7,#8,#9){\begin{Bmatrix}#1&#2&#3\\#4&#5&#6\\#7&#8&#9\end{Bmatrix}}

\begin{document}

\setkeys{Gin}{width=3.25 in}

\title{Selective addressing of high-rank atomic polarization moments}
\author{V. V. Yashchuk}\email{yashchuk@socrates.berkeley.edu}
\affiliation{Department of Physics, University of California at
Berkeley, Berkeley, California 94720-7300}
\author{D. Budker}
\affiliation{Department of Physics, University of California at
Berkeley, Berkeley, California 94720-7300} \affiliation{Nuclear
Science Division, Lawrence Berkeley National Laboratory, Berkeley,
California 94720}
\author{W. Gawlik}
\affiliation{Instytut Fizyki im. M.
Smoluchowskiego, Uniwersytet Jagiello\'{n}ski, Reymonta 4, 30-059
Krakow, Poland}
\author{D. F. Kimball}
\affiliation{Department of Physics, University of California at
Berkeley, Berkeley, California 94720-7300}
\author{Yu. P. Malakyan}
\affiliation{Institute for Physical Research, National Academy of
Sciences of Armenia, Ashtarak-2, 378410, Armenia}
\author{S. M. Rochester}
\affiliation{Department of Physics, University of California at
Berkeley, Berkeley, California 94720-7300}

\date{\today}

\begin{abstract}
We describe a method of selective generation and study of
polarization moments of up to the highest rank $\kappa=2F$
possible for a quantum state with total angular momentum $F$. The
technique is based on nonlinear magneto-optical rotation with
frequency-modulated light. Various polarization moments are
distinguished by the periodicity of light-polarization rotation
induced by the atoms during Larmor precession and exhibit distinct
light-intensity and frequency dependences. We apply the method to
study polarization moments of $^{87}$Rb atoms contained in a vapor
cell with antirelaxation coating. Distinct ultra-narrow (1-Hz
wide) resonances, corresponding to different multipoles, appear in
the magnetic-field dependence of the optical rotation. The use of
the highest-multipole resonances has important applications in
quantum and nonlinear optics and in magnetometry.
\end{abstract}



\maketitle

High-rank polarization moments (PM) and associated high-order
coherences have recently drawn attention (see \cite{Lob96,
Lob97,Sut93,Sut94,Xu97a,Xu97b,Bud2002RMP,Mat2003OL} and references
therein) because they may enhance nonlinear optical effects
important in applications such as electromagnetically induced
transparency \cite{Har97}, creation of nonclassical atomic and
photonic states \cite{Par93}, realization of photon blockades
\cite{Ima97}, quantum gates \cite{Tur95}, photonic switches
\cite{Har98}, and atomic magnetometry \cite{Ale97,Oku2001}.

While signatures of high-order PM were detected in several
experiments \cite{Sut93,Sut94,Xu97a,Xu97b,Mat2003OL}, the methods
used in these investigations are not sufficiently selective and/or
do not allow real-time manipulation of particular multipoles. Here
we describe a method, based on nonlinear optical rotation with
frequency-modulated light (FM NMOR) \cite{Bud2002FM}, by which one
can selectively induce, control, and study any possible multipole
moment. Applying the method to $^{87}$Rb atoms in a
paraffin-coated cell \cite{AlexandrovLPh96,AleLIAD}, we have
verified the expected power and spectral dependences of the
resonant signals and obtained a quantitative comparison of
relaxation rates for the even-rank moments.

\begin{figure}
\includegraphics{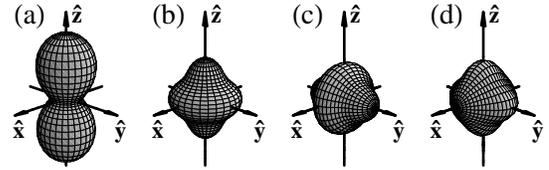}
\caption{Polarization moments visualized as discussed in
\cite{Roc2001} with surfaces for which the distance to the origin
in a given direction is proportional to the probability of finding
the projection $M=F$ along this direction. (a): ``pure" quadrupole
$\kappa=2,q=0$; (b): $\kappa=4,q=0$ hexadecapole; (c): same as in
(b), but rotated by $\pi/2$ around the $x$-axis; (d): the average
of (b) and (c), which has a 4-fold symmetry with respect to
rotations around $\mathbf{\hat{x}}$. In all cases, the minimum
necessary amount of $\rho^{(0)}_0$ was added to ensure that all
sublevel populations are non-negative \cite{Roc2001}. Probability
surfaces (a) and (d) rotating around $\mathbf{\hat{x}}$-directed
magnetic field with the Larmor frequency correspond to the
polarization states produced in this
experiment.}\label{QuadHexFig}
\end{figure}

The density matrix in the $M,M'$ representation for a state with
total angular momentum $F$ can be decomposed into PM of rank
$\kappa=0\ldots 2F$, uncoupled under rotations, with components
$q=-\kappa\ldots\kappa$:
\begin{equation}
\rho_q^{(\kappa)}=\sum_{M,M'=-F}^F(-1)^{F-M'}\cg{F,M,F,-M'}{\kappa,q}\rho_{M,M'},
\label{Eq_MM'to_Kappa_q}\nonumber
\end{equation}
where $\cg{\dots}{\dots}$ indicate the Clebsch-Gordan coefficients
(see, for example, \cite{Var88}). For a given choice of
quantization axis, $|q|\ne 0$ components of the PM are related to
coherences between Zeeman sublevels for which $\Delta M=M-M'=q$,
while $q=0$ components depend on sublevel populations. Thus,
coherences $\rho_{M,M'}$ contribute to all PM with $\kappa \ge
|\Delta M|$ and a $|\Delta M|=2F$ coherence is uniquely associated
with the highest PM for a given state, e.g., the quadrupole moment
($\kappa=2$) for $F=1$, or the hexadecapole ($\kappa=4$) for
$F=2$. The method introduced here exploits the different axial
symmetries of the PM (2-fold and 4-fold for the quadrupole and
hexadecapole, respectively; Fig.\ \ref{QuadHexFig}) to selectively
create and detect them (see also \cite{Sut93,Sut94,Xu97a,Xu97b}).

\begin{figure}
\includegraphics[width=3 in]{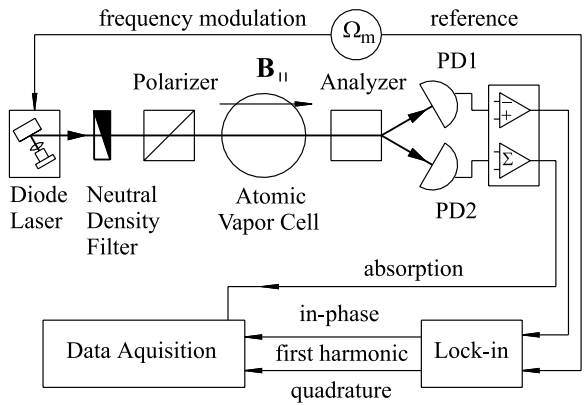}
\caption{Simplified diagram of the FM NMOR setup. The balanced
polarimeter incorporating the analyzer and photodiodes PD1 and PD2
detects signals due to time-dependent optical rotation of linearly
polarized frequency-modulated light.}\label{FM_NMOR_Apparatus_Fig}
\end{figure}

While multipole moments of rank $\kappa\le 2$ can be easily
generated and detected with weak light (since a photon has spin
one), higher-rank moments require multi-photon interactions for
both production \emph{and detection}. In the present method, we
use a single laser beam (which is still of sub-mW power) for the
nonlinear interactions required to pump and probe the
high-multipole moments.

\begin{figure}
\includegraphics[width=3 in]{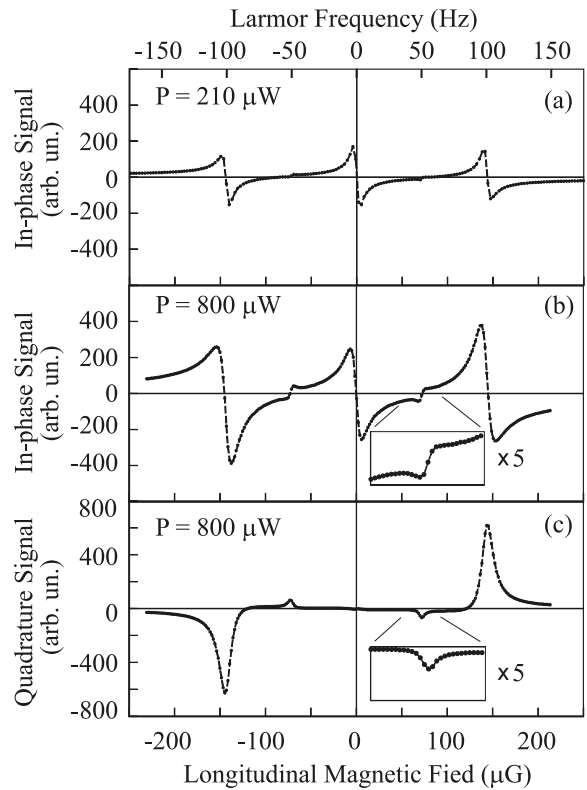}
\caption{An example of the magnetic-field dependence of the FM
NMOR signals showing quadrupole resonances at $B=\pm 143.0\ \mu$G,
and the hexadecapole resonances at $\pm 71.5\ \mu$G. Laser
modulation frequency is 200$\ $Hz, modulation amplitude is 40$\
$MHz peak-to-peak; the central frequency is tuned to the
low-frequency slope of the $F=2\rightarrow F'=1$ absorption line.
Plots (a,b) show the in-phase component of the signal at two
different light powers; plot (c) shows the quadrature component.
Note the increase in the relative size of the hexadecapole signals
at the higher power. The insets show zooms on hexadecapole
resonances.}\label{BxDependence}
\end{figure}

Under conditions of our experiment, FM NMOR (Fig.$\
$\ref{FM_NMOR_Apparatus_Fig}) can be understood as a three-stage
(pump, precession, probe) process: atoms are polarized in an
interaction with the laser beam (whose diameter is much smaller
than the vapor cell dimensions), then leave the beam and bounce
around the cell while undergoing Larmor precession, and finally
return into the laser beam region and undergo the ``probe"
interaction. The laser light is frequency modulated, causing the
optical pumping and probing to acquire a periodic time dependence.
When the pumping rate is synchronized with the precession of
atomic polarization, a resonance occurs and the atomic medium is
pumped into a polarized state which rotates around the direction
of the magnetic field at the Larmor frequency $\Omega_L$. The
optical properties of the medium are modulated at the frequency
$\kappa\Omega_L$, due to the symmetry of atomic polarization with
rank $\kappa$. For example, for the quadrupole moment the
modulation is at $2\Omega_L$, and for the hexadecapole it is at
$4\Omega_L$ [Fig. \ref{QuadHexFig}(a,d)]. This periodic change of
the optical properties of the atomic vapor modulates the angle of
the light polarization, leading to the FM NMOR resonances. If the
time-dependent optical rotation is measured at the first harmonic
of $\Omega_m$, a resonance is seen when $\Omega_m$ coincides with
$\kappa\Omega_L$ (Fig. \ref{BxDependence}) \footnote{Additional
resonances can be observed at higher harmonics \cite{Bud2002FM}.}.

At the resonance for a PM of rank $\kappa$ (which should be absent
for states with $2F\le \kappa$) we expect the signal amplitude to
go as the $\kappa$-th power of light intensity at low intensities,
as discussed below. Next, we show that these predictions are
verified in this experiment.

\begin{figure}
\includegraphics[width=3 in]{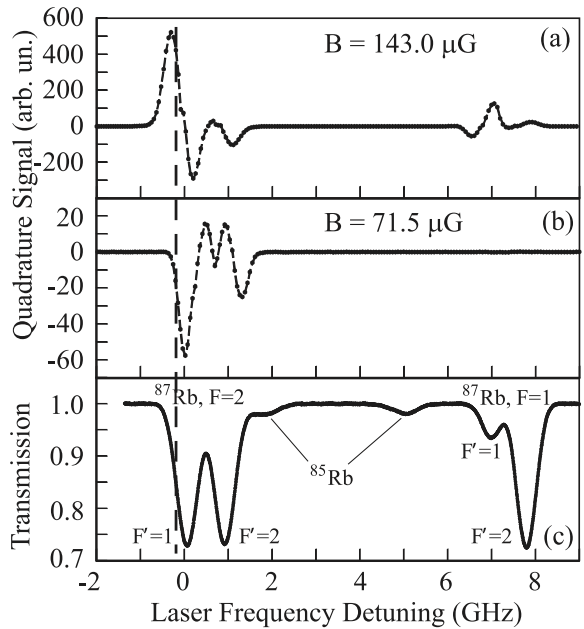}
\caption{Spectral dependences of the quadrupole (a) and the
hexadecapole (b) signals measured at light power of $800\ \mu$W,
and the low-power ($.4\ \mu$W) transmission spectrum (no FM) (c).
Note the different spectral dependences for the FM NMOR signals in
(a) and (b), and in particular, the absence of the signal at the
$F=1\rightarrow F'$ transitions in (b). The vertical line
indicates the central laser frequency where the measurements
represented in Figs. \ref{BxDependence}, \ref{AmpVsInt_Fig}, and
\ref{GamVsInt_Fig} were taken.}\label{FigFrDepen2}
\end{figure}

In the present experiment (Fig. \ref{FM_NMOR_Apparatus_Fig}), we
used the FM NMOR technique \cite{Bud2002FM} with $^{87}$Rb atoms.
The central laser frequency was tuned near various hfs components
of the D1 line. The typical light power was a few hundred $\mu$W
and the laser beam diameter was $\sim$$3\ $mm. The laser frequency
was modulated at $\Omega_m/(2\pi)$ from $50\ $Hz to 1$\ $kHz, and
the frequency modulation amplitude was approximately $40\ $MHz
(peak-to-peak). The vapor cell, with isotopically enriched
$^{87}$Rb, is 10$\ $cm in diameter and has an antirelaxation
coating and no buffer gas. The cell is surrounded with four layers
of magnetic shielding. A system of coils inside the innermost
shield is used to compensate the residual fields (at a level
$\lesssim 0.1\ \mu$G) and first-order gradients, and to apply a
well-controlled, arbitrarily directed magnetic field to the atoms.
This allows observation of FM NMOR resonances with magnetic-field
widths of about 1$\ \mu$G in the low-light-intensity limit.

Figure \ref{BxDependence} shows the magnetic-field dependence of
the observed FM NMOR signals. The central laser frequency was
tuned to the low-frequency slope of the $F=2\rightarrow F'=1$
absorption line as shown in Fig. \ref{FigFrDepen2}. At relatively
low light power [Fig. \ref{BxDependence}(a)], there are three
prominent resonances: one at $B=0$, and two corresponding to
$2\Omega_L=\Omega_m$. Much smaller signals, whose relative
amplitudes rapidly grow with light power, are seen at
$4\Omega_L=\Omega_m$, the expected positions of the hexadecapole
resonances.

The spectral dependences for both types of resonance signals with
fixed magnetic field and modulation frequency are shown in Fig.
\ref{FigFrDepen2}. While the quadrupole resonance signals
[$\Omega_m=2\Omega_L$; Fig. \ref{FigFrDepen2}(a)] are observed for
both ground-state hyperfine components, no signals are observed
for the hexadecapole resonances [$\Omega_m=4\Omega_L$; Fig.\
\ref{FigFrDepen2}(b)] near the lines involving the $F=1$ ground
state, which can not support a hexadecapole moment \footnote{The
spectral dependences shown in Fig. \ref{FigFrDepen2}(a,b) were
obtained by subtracting the average of spectra of the quadrature
component taken approximately $15\ \mu$G above and below the
resonance from the spectrum recorded with magnetic field set at
the resonance value. This procedure renders a measurement free
from the small residual contribution from the zero-field resonance
arising from the ``transit" effect \cite{Bud2002FM}.}. The
different shapes of the quadrupole and hexadecapole spectra near
the $F=2$ transition group require further analysis.

\begin{figure}
\includegraphics[width=3 in]{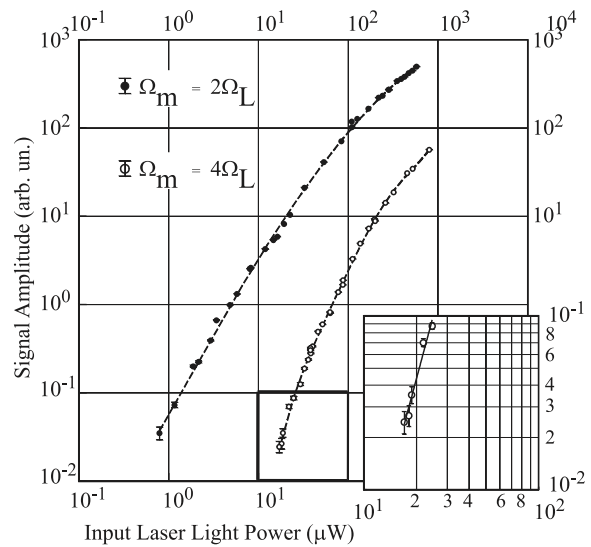}
\caption{Signal amplitude vs. the input laser power. Filled
circles: the quadrupole resonance; open circles: the hexadecapole
resonance. The inset shows the expanded low-power region for the
latter. From these data, we determine the initial linear slope on
these log-log plots (corresponding to the exponent in the power
dependence of the signal) as $1.96(6)$ and $3.75(37)$ for the
quadrupole and hexadecapole cases, respectively. The corresponding
expected exponents are 2 and 4.}\label{AmpVsInt_Fig}
\end{figure}

Light-intensity dependences of the resonance amplitudes are shown
in Fig. \ref{AmpVsInt_Fig}. The observed low-intensity ($I$)
asymptotics of these curves (which show saturation at higher
intensities) scale approximately as $I^2$ and $I^4$. These power
dependences and many other salient features of the observed
resonances can be understood with the help of a simple model of an
$F=2\rightarrow F'=1$ transition with independent pumping,
evolution in a magnetic field, and probing, in which to
lowest-order, the quadrupole and hexadecapole moments are first
($\rho^{(2)}\propto I_{pump}$) and second order
($\rho^{(4)}\propto I_{pump}^2$) in light intensity, respectively.
The component of the signal due to the induced optical rotation in
the probe beam proportional to the quadrupole moment goes as
$I_{probe}\rho^{(2)}$, while that proportional to the hexadecapole
moment goes as $I_{probe}^2\rho^{(4)}$. The intensity dependences
predicted by this model, as well as the periodicity of the signals
with respect to Larmor precession and the relative widths of the
resonances, match the observations from the experiment, in which a
single beam serves as pump and probe.

\begin{figure}
\includegraphics[width=3 in]{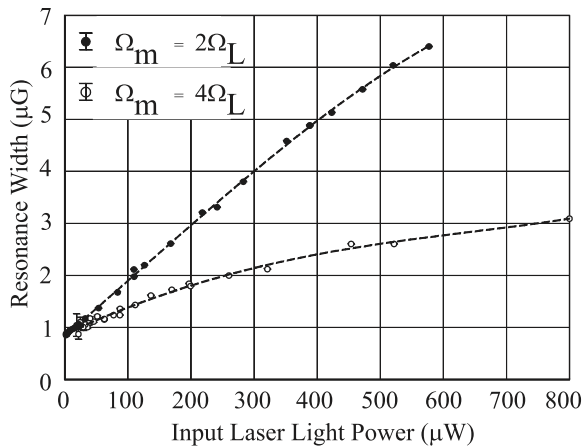}
\caption{Resonance width vs. the input laser power. Filled
circles: the quadrupole resonance; open circles: the hexadecapole
resonance. The widths extrapolated to zero light power are found
to be $\Delta B_{q}=0.848(4)\ \mu$G and $\Delta B_{h}=0.904(33)\
\mu$G for the quadrupole and hexadecapole cases,
respectively.}\label{GamVsInt_Fig}
\end{figure}

Figure \ref{GamVsInt_Fig} shows the dependence of the resonance
widths on power at a fixed magnetic field. While both the
quadrupole and hexadecapole resonances exhibit power broadening,
it is much less pronounced in the latter case. This is important
for applications such as magnetometry because it allows operation
at higher light powers with better statistical sensitivity
\cite{Bud2000Sens}. In the zero-power limit, we find that the
resonance widths (Fig. \ref{GamVsInt_Fig}) for the two types of
resonances tend to values near $1\ \mu$G with ratio $\Delta
B_{q}/\Delta B_{h}=0.94(4)$. The expected ratio of these widths is
twice the ratio of the light-independent relaxation rates for the
PM with $\kappa=2$ and 4 \footnote{It can be shown that the width
of an FM NMOR resonance corresponding to a PM of rank $\kappa$
goes as $\gamma_\kappa/(\kappa g \mu)$, where $\gamma_\kappa$ is
the rate of light-independent relaxation for a PM of rank
$\kappa$, $g$ is the gyromagnetic ratio, and $\mu$ is the Bohr
magneton.}. Thus, we find that the light-independent relaxation
rate for the hexadecapole is approximately twice that of the
quadrupole. Relaxation of the PM in our experimental conditions is
dominated by the residual relaxation on the paraffin-coated cell
walls and spin-exchange collisions between Rb atoms. The
electron-randomization collision model (see, for example,
\cite{Kni88}), predicts that the quadrupole moment relaxes at a
rate $3/8$ that of the hexadecapole moment, which relaxes at the
electron-randomization rate. Thus, the observed ratio of the
widths is close to the expected.

In conclusion, we have developed and applied a new technique for
the study of high atomic-polarization moments, allowing their
selective creation and detection via corresponding well-resolved
resonances. The study of the highest-rank PM signals in FM NMOR is
important for the applications of this technique in Earth-field
magnetometry \cite{Bud2002FM}. This is because energy separation
for Zeeman sublevels with $\Delta M=2F$ of the $F=I+1/2$ states of
the alkali atoms is linear in the magnetic field, while
separations between other sublevels are generally nonlinear due to
magnetic-field mixing of states of different $F$. Therefore, it is
advantageous to use of the FM NMOR resonances due to the $\Delta
M=2F$ coherence, as will be discussed in detail elsewhere.

Future work will explore the extension of the technique to
separated pump and probe light beams and application to higher
angular-momentum states. This will allow further optimization of
the selective control of multipole moments and will elucidate the
exact mechanisms responsible for the nonlinear light-atom
interactions involved at the pump and the probe stages.
Specifically, we will explore the role of conversion of the
high-rank moments into alignment and orientation under the
combined action of the magnetic field and the light shifts, which
can be of crucial importance in nonlinear optical rotation
\cite{Bud2000AOC,Roc2001,Roc2001SR,Bud2002RMP}.

We thank M.~Zolotorev for useful discussions and A.~Vaynberg for
crafting parts of the apparatus. This work has been supported by
the Office of Naval Research (grant N00014-97-1-0214); by NRC
US-Poland Twinning grant 015369, and by the Polish Committee for
Scientific Research (grant PBZ/KBN/043/PO3/2001) and the National
Laboratory of AMO Physics in Toru\'{n}, Poland; by a US-Armenian
bilateral  Grant CRDF AP2-3213/NFSAT PH 071-02; by NSF; and by the
Director, Office of Science, Nuclear Science Division, of the U.S.
Department of Energy under contract DE-AC03-76SF00098. D.B. also
acknowledges the support of the Miller Institute for Basic
Research in Science.

\bibliography{NMObibl}

\begin{thebibliography}{24}
\expandafter\ifx\csname natexlab\endcsname\relax\def\natexlab#1{#1}\fi
\expandafter\ifx\csname bibnamefont\endcsname\relax
  \def\bibnamefont#1{#1}\fi
\expandafter\ifx\csname bibfnamefont\endcsname\relax
  \def\bibfnamefont#1{#1}\fi
\expandafter\ifx\csname citenamefont\endcsname\relax
  \def\citenamefont#1{#1}\fi
\expandafter\ifx\csname url\endcsname\relax
  \def\url#1{\texttt{#1}}\fi
\expandafter\ifx\csname urlprefix\endcsname\relax\def\urlprefix{URL }\fi
\providecommand{\bibinfo}[2]{#2}
\providecommand{\eprint}[2][]{\url{#2}}

\bibitem[{\citenamefont{\L{}obodzi\'{n}ski and Gawlik}(1996)}]{Lob96}
\bibinfo{author}{\bibfnamefont{B.}~\bibnamefont{\L{}obodzi\'{n}ski}}
  \bibnamefont{and} \bibinfo{author}{\bibfnamefont{W.}~\bibnamefont{Gawlik}},
  \bibinfo{journal}{Phys. Rev. A}
  \textbf{\bibinfo{volume}{54}}(\bibinfo{number}{3}), \bibinfo{pages}{2238}
  (\bibinfo{year}{1996}).

\bibitem[{\citenamefont{\L{}obodzi\'{n}ski and Gawlik}(1997)}]{Lob97}
\bibinfo{author}{\bibfnamefont{B.}~\bibnamefont{\L{}obodzi\'{n}ski}}
  \bibnamefont{and} \bibinfo{author}{\bibfnamefont{W.}~\bibnamefont{Gawlik}},
  \bibinfo{journal}{Phys. Scr.} \textbf{\bibinfo{volume}{T70}},
  \bibinfo{pages}{138} (\bibinfo{year}{1997}).

\bibitem[{\citenamefont{Suter et~al.}(1993)\citenamefont{Suter, Marty, and
  Klepel}}]{Sut93}
\bibinfo{author}{\bibfnamefont{D.}~\bibnamefont{Suter}},
  \bibinfo{author}{\bibfnamefont{T.}~\bibnamefont{Marty}}, \bibnamefont{and}
  \bibinfo{author}{\bibfnamefont{H.}~\bibnamefont{Klepel}},
  \bibinfo{journal}{Opt. Lett.}
  \textbf{\bibinfo{volume}{18}}(\bibinfo{number}{7}), \bibinfo{pages}{531}
  (\bibinfo{year}{1993}).

\bibitem[{\citenamefont{Suter and Marty}(1994)}]{Sut94}
\bibinfo{author}{\bibfnamefont{D.}~\bibnamefont{Suter}} \bibnamefont{and}
  \bibinfo{author}{\bibfnamefont{T.}~\bibnamefont{Marty}}, \bibinfo{journal}{J.
  Opt. Soc. Am. B} \textbf{\bibinfo{volume}{11}}(\bibinfo{number}{1}),
  \bibinfo{pages}{242} (\bibinfo{year}{1994}).

\bibitem[{\citenamefont{Xu et~al.}(1997{\natexlab{a}})\citenamefont{Xu,
  W\"{a}ckerle, and Mehring}}]{Xu97a}
\bibinfo{author}{\bibfnamefont{J.~D.} \bibnamefont{Xu}},
  \bibinfo{author}{\bibfnamefont{G.}~\bibnamefont{W\"{a}ckerle}},
  \bibnamefont{and} \bibinfo{author}{\bibfnamefont{M.}~\bibnamefont{Mehring}},
  \bibinfo{journal}{Phys. Rev. A}
  \textbf{\bibinfo{volume}{55}}(\bibinfo{number}{1}), \bibinfo{pages}{206}
  (\bibinfo{year}{1997}{\natexlab{a}}).

\bibitem[{\citenamefont{Xu et~al.}(1997{\natexlab{b}})\citenamefont{Xu,
  W\"{a}ckerle, and Mehring}}]{Xu97b}
\bibinfo{author}{\bibfnamefont{J.~D.} \bibnamefont{Xu}},
  \bibinfo{author}{\bibfnamefont{G.}~\bibnamefont{W\"{a}ckerle}},
  \bibnamefont{and} \bibinfo{author}{\bibfnamefont{M.}~\bibnamefont{Mehring}},
  \bibinfo{journal}{Z. Phys. D}
  \textbf{\bibinfo{volume}{42}}(\bibinfo{number}{1}), \bibinfo{pages}{5}
  (\bibinfo{year}{1997}{\natexlab{b}}).

\bibitem[{\citenamefont{Budker et~al.}(2002{\natexlab{a}})\citenamefont{Budker,
  Gawlik, Kimball, Rochester, Yashchuk, and Weis}}]{Bud2002RMP}
\bibinfo{author}{\bibfnamefont{D.}~\bibnamefont{Budker}},
  \bibinfo{author}{\bibfnamefont{W.}~\bibnamefont{Gawlik}},
  \bibinfo{author}{\bibfnamefont{D.~F.} \bibnamefont{Kimball}},
  \bibinfo{author}{\bibfnamefont{S.~M.} \bibnamefont{Rochester}},
  \bibinfo{author}{\bibfnamefont{V.~V.} \bibnamefont{Yashchuk}},
  \bibnamefont{and} \bibinfo{author}{\bibfnamefont{A.}~\bibnamefont{Weis}},
  \bibinfo{journal}{Rev. Mod. Phys.}
  \textbf{\bibinfo{volume}{74}}(\bibinfo{number}{4}), \bibinfo{pages}{1153}
  (\bibinfo{year}{2002}{\natexlab{a}}).

\bibitem[{\citenamefont{Matsko et~al.}(2003)\citenamefont{Matsko, Novikova,
  Welch, and Zubairy}}]{Mat2003OL}
\bibinfo{author}{\bibfnamefont{A.~B.} \bibnamefont{Matsko}},
  \bibinfo{author}{\bibfnamefont{I.}~\bibnamefont{Novikova}},
  \bibinfo{author}{\bibfnamefont{G.~R.} \bibnamefont{Welch}}, \bibnamefont{and}
  \bibinfo{author}{\bibfnamefont{M.~S.} \bibnamefont{Zubairy}},
  \bibinfo{journal}{Opt. Lett.}
  \textbf{\bibinfo{volume}{28}}(\bibinfo{number}{2}), \bibinfo{pages}{96}
  (\bibinfo{year}{2003}).

\bibitem[{\citenamefont{Harris}(1997)}]{Har97}
\bibinfo{author}{\bibfnamefont{S.~E.} \bibnamefont{Harris}},
  \bibinfo{journal}{Phys. Today}
  \textbf{\bibinfo{volume}{50}}(\bibinfo{number}{7}), \bibinfo{pages}{36}
  (\bibinfo{year}{1997}).

\bibitem[{\citenamefont{Parkins et~al.}(1993)\citenamefont{Parkins, Marte,
  Zoller, and Kimble}}]{Par93}
\bibinfo{author}{\bibfnamefont{A.~S.} \bibnamefont{Parkins}},
  \bibinfo{author}{\bibfnamefont{P.}~\bibnamefont{Marte}},
  \bibinfo{author}{\bibfnamefont{P.}~\bibnamefont{Zoller}}, \bibnamefont{and}
  \bibinfo{author}{\bibfnamefont{H.~J.} \bibnamefont{Kimble}},
  \bibinfo{journal}{Phys. Rev. Lett.}
  \textbf{\bibinfo{volume}{71}}(\bibinfo{number}{19}), \bibinfo{pages}{3095}
  (\bibinfo{year}{1993}).

\bibitem[{\citenamefont{Imamoglu et~al.}(1997)\citenamefont{Imamoglu, Schmidt,
  Woods, and Deutsch}}]{Ima97}
\bibinfo{author}{\bibfnamefont{A.}~\bibnamefont{Imamoglu}},
  \bibinfo{author}{\bibfnamefont{H.}~\bibnamefont{Schmidt}},
  \bibinfo{author}{\bibfnamefont{G.}~\bibnamefont{Woods}}, \bibnamefont{and}
  \bibinfo{author}{\bibfnamefont{M.}~\bibnamefont{Deutsch}},
  \bibinfo{journal}{Phys. Rev. Lett.}
  \textbf{\bibinfo{volume}{79}}(\bibinfo{number}{8}), \bibinfo{pages}{1467}
  (\bibinfo{year}{1997}).

\bibitem[{\citenamefont{Turchette et~al.}(1995)\citenamefont{Turchette, Hood,
  Lange, Mabuchi, and Kimble}}]{Tur95}
\bibinfo{author}{\bibfnamefont{Q.~A.} \bibnamefont{Turchette}},
  \bibinfo{author}{\bibfnamefont{C.~J.} \bibnamefont{Hood}},
  \bibinfo{author}{\bibfnamefont{W.}~\bibnamefont{Lange}},
  \bibinfo{author}{\bibfnamefont{H.}~\bibnamefont{Mabuchi}}, \bibnamefont{and}
  \bibinfo{author}{\bibfnamefont{H.~J.} \bibnamefont{Kimble}},
  \bibinfo{journal}{Phys. Rev. Lett.}
  \textbf{\bibinfo{volume}{75}}(\bibinfo{number}{25}), \bibinfo{pages}{4710}
  (\bibinfo{year}{1995}).

\bibitem[{\citenamefont{Harris and Yamamoto}(1998)}]{Har98}
\bibinfo{author}{\bibfnamefont{S.~E.} \bibnamefont{Harris}} \bibnamefont{and}
  \bibinfo{author}{\bibfnamefont{Y.}~\bibnamefont{Yamamoto}},
  \bibinfo{journal}{Phys. Rev. Lett.}
  \textbf{\bibinfo{volume}{81}}(\bibinfo{number}{17}), \bibinfo{pages}{3611}
  (\bibinfo{year}{1998}).

\bibitem[{\citenamefont{Alexandrov et~al.}(1997)\citenamefont{Alexandrov,
  Pazgalev, and Rasson}}]{Ale97}
\bibinfo{author}{\bibfnamefont{E.~B.} \bibnamefont{Alexandrov}},
  \bibinfo{author}{\bibfnamefont{A.~S.} \bibnamefont{Pazgalev}},
  \bibnamefont{and} \bibinfo{author}{\bibfnamefont{J.~L.}
  \bibnamefont{Rasson}}, \bibinfo{journal}{Opt. Spectrosk.}
  \textbf{\bibinfo{volume}{82}}(\bibinfo{number}{1}), \bibinfo{pages}{14}
  (\bibinfo{year}{1997}).

\bibitem[{\citenamefont{Okunevich}(2001)}]{Oku2001}
\bibinfo{author}{\bibfnamefont{A.~I.} \bibnamefont{Okunevich}},
  \bibinfo{journal}{Opt. Spectrosk.}
  \textbf{\bibinfo{volume}{91}}(\bibinfo{number}{2}), \bibinfo{pages}{193}
  (\bibinfo{year}{2001}).

\bibitem[{\citenamefont{Budker et~al.}(2002{\natexlab{b}})\citenamefont{Budker,
  Kimball, Yashchuk, and Zolotorev}}]{Bud2002FM}
\bibinfo{author}{\bibfnamefont{D.}~\bibnamefont{Budker}},
  \bibinfo{author}{\bibfnamefont{D.~F.} \bibnamefont{Kimball}},
  \bibinfo{author}{\bibfnamefont{V.~V.} \bibnamefont{Yashchuk}},
  \bibnamefont{and}
  \bibinfo{author}{\bibfnamefont{M.}~\bibnamefont{Zolotorev}},
  \bibinfo{journal}{Phys. Rev. A} \textbf{\bibinfo{volume}{65}},
  \bibinfo{pages}{055403} (\bibinfo{year}{2002}{\natexlab{b}}).

\bibitem[{\citenamefont{Alexandrov et~al.}(1996)\citenamefont{Alexandrov,
  Balabas, Pasgalev, Vershovskii, and Yakobson}}]{AlexandrovLPh96}
\bibinfo{author}{\bibfnamefont{E.~B.} \bibnamefont{Alexandrov}},
  \bibinfo{author}{\bibfnamefont{M.~V.} \bibnamefont{Balabas}},
  \bibinfo{author}{\bibfnamefont{A.~S.} \bibnamefont{Pasgalev}},
  \bibinfo{author}{\bibfnamefont{A.~K.} \bibnamefont{Vershovskii}},
  \bibnamefont{and} \bibinfo{author}{\bibfnamefont{N.~N.}
  \bibnamefont{Yakobson}}, \bibinfo{journal}{Laser Physics}
  \textbf{\bibinfo{volume}{6}}(\bibinfo{number}{2}), \bibinfo{pages}{244}
  (\bibinfo{year}{1996}).

\bibitem[{\citenamefont{Alexandrov et~al.}(2002)\citenamefont{Alexandrov,
  Balabas, Budker, English, Kimball, Li, and Yashchuk}}]{AleLIAD}
\bibinfo{author}{\bibfnamefont{E.~B.} \bibnamefont{Alexandrov}},
  \bibinfo{author}{\bibfnamefont{M.~V.} \bibnamefont{Balabas}},
  \bibinfo{author}{\bibfnamefont{D.}~\bibnamefont{Budker}},
  \bibinfo{author}{\bibfnamefont{D.~S.} \bibnamefont{English}},
  \bibinfo{author}{\bibfnamefont{D.~F.} \bibnamefont{Kimball}},
  \bibinfo{author}{\bibfnamefont{C.~H.} \bibnamefont{Li}}, \bibnamefont{and}
  \bibinfo{author}{\bibfnamefont{V.}~\bibnamefont{Yashchuk}},
  \bibinfo{journal}{Phys. Rev. A}
  \textbf{\bibinfo{volume}{66}}(\bibinfo{number}{4}) (\bibinfo{year}{2002}).

\bibitem[{\citenamefont{Varshalovich et~al.}(1988)\citenamefont{Varshalovich,
  Moskalev, and Khersonskii}}]{Var88}
\bibinfo{author}{\bibfnamefont{D.~A.} \bibnamefont{Varshalovich}},
  \bibinfo{author}{\bibfnamefont{A.~N.} \bibnamefont{Moskalev}},
  \bibnamefont{and} \bibinfo{author}{\bibfnamefont{V.~K.}
  \bibnamefont{Khersonskii}}, \emph{\bibinfo{title}{Quantum theory of angular
  momentum: irreducible tensors, spherical harmonics, vector coupling
  coefficients, 3nj symbols}} (\bibinfo{publisher}{World Scientific},
  \bibinfo{address}{Singapore}, \bibinfo{year}{1988}).

\bibitem[{\citenamefont{Rochester and Budker}(2001)}]{Roc2001}
\bibinfo{author}{\bibfnamefont{S.~M.} \bibnamefont{Rochester}}
  \bibnamefont{and} \bibinfo{author}{\bibfnamefont{D.}~\bibnamefont{Budker}},
  \bibinfo{journal}{Am. J. Phys.}
  \textbf{\bibinfo{volume}{69}}(\bibinfo{number}{4}), \bibinfo{pages}{450}
  (\bibinfo{year}{2001}).

\bibitem[{\citenamefont{Budker et~al.}(2000{\natexlab{a}})\citenamefont{Budker,
  Kimball, Rochester, Yashchuk, and Zolotorev}}]{Bud2000Sens}
\bibinfo{author}{\bibfnamefont{D.}~\bibnamefont{Budker}},
  \bibinfo{author}{\bibfnamefont{D.~F.} \bibnamefont{Kimball}},
  \bibinfo{author}{\bibfnamefont{S.~M.} \bibnamefont{Rochester}},
  \bibinfo{author}{\bibfnamefont{V.~V.} \bibnamefont{Yashchuk}},
  \bibnamefont{and}
  \bibinfo{author}{\bibfnamefont{M.}~\bibnamefont{Zolotorev}},
  \bibinfo{journal}{Phys. Rev. A}
  \textbf{\bibinfo{volume}{62}}(\bibinfo{number}{4}), \bibinfo{pages}{043403/1}
  (\bibinfo{year}{2000}{\natexlab{a}}).

\bibitem[{\citenamefont{Knize et~al.}(1988)\citenamefont{Knize, Wu, and
  Happer}}]{Kni88}
\bibinfo{author}{\bibfnamefont{R.~J.} \bibnamefont{Knize}},
  \bibinfo{author}{\bibfnamefont{Z.}~\bibnamefont{Wu}}, \bibnamefont{and}
  \bibinfo{author}{\bibfnamefont{W.}~\bibnamefont{Happer}},
  \bibinfo{journal}{Advances in Atomic and Molecular Physics}
  \textbf{\bibinfo{volume}{24}}, \bibinfo{pages}{223} (\bibinfo{year}{1988}).

\bibitem[{\citenamefont{Budker et~al.}(2000{\natexlab{b}})\citenamefont{Budker,
  Kimball, Rochester, and Yashchuk}}]{Bud2000AOC}
\bibinfo{author}{\bibfnamefont{D.}~\bibnamefont{Budker}},
  \bibinfo{author}{\bibfnamefont{D.~F.} \bibnamefont{Kimball}},
  \bibinfo{author}{\bibfnamefont{S.~M.} \bibnamefont{Rochester}},
  \bibnamefont{and} \bibinfo{author}{\bibfnamefont{V.~V.}
  \bibnamefont{Yashchuk}}, \bibinfo{journal}{Phys. Rev. Lett.}
  \textbf{\bibinfo{volume}{85}}(\bibinfo{number}{10}), \bibinfo{pages}{2088}
  (\bibinfo{year}{2000}{\natexlab{b}}).

\bibitem[{\citenamefont{Rochester et~al.}(2001)\citenamefont{Rochester, Hsiung,
  Budker, Chiao, Kimball, and Yashchuk}}]{Roc2001SR}
\bibinfo{author}{\bibfnamefont{S.~M.} \bibnamefont{Rochester}},
  \bibinfo{author}{\bibfnamefont{D.~S.} \bibnamefont{Hsiung}},
  \bibinfo{author}{\bibfnamefont{D.}~\bibnamefont{Budker}},
  \bibinfo{author}{\bibfnamefont{R.~Y.} \bibnamefont{Chiao}},
  \bibinfo{author}{\bibfnamefont{D.~F.} \bibnamefont{Kimball}},
  \bibnamefont{and} \bibinfo{author}{\bibfnamefont{V.~V.}
  \bibnamefont{Yashchuk}}, \bibinfo{journal}{Phys. Rev. A}
  \textbf{\bibinfo{volume}{63}}(\bibinfo{number}{4}), \bibinfo{pages}{043814/1}
  (\bibinfo{year}{2001}).

\end{thebibliography}

\end{document}